# Pattern Atlas

Frank Leymann[0000-0002-9123-259X] and Johanna Barzen[0000-0001-8397-7973]

IAAS, University of Stuttgart, Universitätsstr. 38, 70569 Stuttgart, Germany
`{firstname.lastname}@iaas.uni-stuttgart.de`

**Abstract.** Pattern languages are well-established in the software architecture community. Many different aspects of creating a software architecture are addressed by such languages. Thus, several pattern languages have to be considered when building a particular architecture. But these pattern languages are isolated, i.e. it is hard to determine the relevant patterns to be applied from the different pattern languages. Moreover, the sum of patterns from different languages may be huge, i.e. restriction to relevant patterns is desirable. In this contribution we envision an encompassing tool, the pattern atlas, that supports building complex systems based on pattern languages. The analogy to cartography motivates the name of the tool.

## 1. Principles of Pattern Languages

Pattern languages have their origin in architecting houses and urban planning [1]. In the meantime it is used in a whole spectrum of fields reaching from information technology (e.g. [5, 11, 12, 13, 14, 16, 18]) to the humanities (e.g. [2, 3, 4]). In this section we will sketch the basics behind pattern languages and their principle use.

### 1.1. The Notion of a Pattern Language

A *pattern* is a proven solution of a recurring problem. "Proven" means that the outlined solution has been successfully applied several times, and the situation in which the solution has been applied was not always the same but showed some variance. This indicates that the corresponding problem occurred more than once (and will occur in future again), i.e. it is "recurring". If it would not be recurring, the effort to document the solution would not be worth spending.

The presented "solution" is generic in the sense that it captures the essence of each of the working solutions in the corresponding contexts but no specific details of the working solution at all. I.e. a pattern is in fact derived by abstraction from the working solutions ($\sigma_1,\ldots, \sigma_n$ in Fig. 1) [10, 16]. Because a pattern's solution is generic it can be applied in new, unforeseen situations. Vice versa, instead of forgetting the working solutions a pattern has been derived from, a pattern may refer to these working solutions, and even future implementations of the abstract solution of the pattern (called "concretizations" in Fig.1) may also be associated [6]. This way, a pattern becomes a source of reuse of working solutions freeing users to implement the abstract solution over and over again.

Often, when facing a certain problem, other problems will occur too. This situation is captured by directed links that point from a pattern to the related other patterns. These links may have various semantics like that the problem of the target pattern often appears jointly with the problem of the source pattern or the solution of the target pattern excludes the solution of the source pattern, respectively. $\Lambda 1$ and $\Lambda 2$ in Fig.1 indicate such links between patterns. Together, patterns form a weighted directed graph the nodes of which are patterns, the edges of which are these links, and the weights are the semantics of the links [7]. Such a graph is referred to as a *pattern language.*

The term "language" indicates the generative nature of a pattern language: by navigating from one pattern to another a corresponding sequence of generic solutions is generated, and these solutions are applied to solve a composite, more complex problem. Such a sequence is referred to as a *solution path* [25].

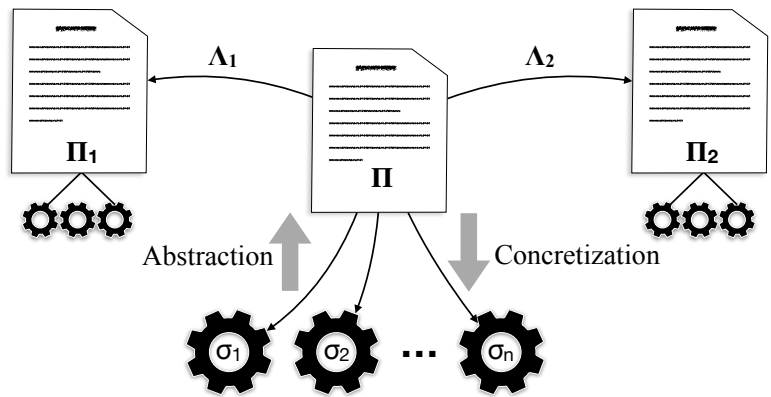

**Fig. 1**. A pattern as an abstraction of working solutions, and related to other patterns.

## 1.2. Using Multiple Pattern Languages

A pattern language is specific for a certain domain, e.g. it addresses solutions of integration problems, cloud computing problems etc. In general, a complex system requires to solve many problems from multiple domains: for example, components have to be realized as microservices, these services must be integrated, they have to be secured, their robustness must be guaranteed - and each problem domain is addressed by a separate pattern language.

In order to solve such complex multi-domain problems, a user will have to navigate across pattern languages. For this purpose, *links* between pattern languages are needed (dashed arrows in Fig. 2) [7]. But the authors of individual pattern languages are most often only concerned with links within pattern languages, e.g. simply because they can't consider a whole plethora of other pattern languages. A further complication is that pattern languages are often published as books, i.e. links across books can't be established, they can't be foreseen because the collection of pattern languages evolve over time. Thus, links between pattern languages have to be realized as separate artifacts.

This is why our existing approach to pattern languages (PatternPedia - see section 2.1) publishes pattern languages as online resources (sometimes in addition to published books) like [20, 21, 22]. Patterns are realized as marked-up documents (e.g. HTML documents), and links between pattern documents are URLs (e.g. hyperlinks). These links are not embedded in the pattern documents but external to the documents: they are directed, pointing from a source document to a target document.

Especially, a link can originate from a document in one pattern language and target a document in another pattern language: this way, cross pattern language links can be established, and they can be established at a later point in time, e.g when a new pattern language appears.

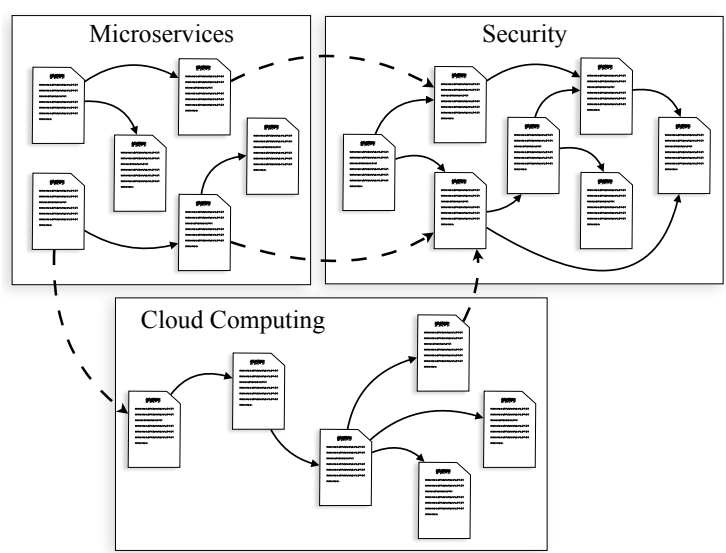

**Fig. 2**. Connecting pattern languages to solve composite problems.

Also, a cross pattern language link may have the semantics that the target pattern concretizes the solution sketched by the source pattern. For example, a cloud computing pattern describes when and how an elastic load balancer is used, and links may point to patterns that show how this is realized in Amazon Web Services or Azure. Thus, a concretization of a pattern is not necessarily an executable, but may be another, refining pattern.

If concretizations are executables, they may be used for automatically building working *solutions* of a composite problem [9]. When deriving a solution path by navigating through a pattern language and selecting working solutions of each of the patterns along this path, proper annotations associated with working solutions help to derive a working solution of the composite problem represented by the solution path [8].

A pattern language might be quite comprehensive, and if multiple pattern languages have to be consulted, the body of knowledge to consider can easily become hardly comprehensible. In such a situation, a pattern language *view* is quite useful [24]: it consists of a subset of relevant patterns from the pattern languages to be considered together with the relevant links within the pattern languages or across pattern languages (see Fig. 3). Such a view is created by a specialist understanding the collection of relevant pattern languages and its applicability to solve cross-domain problems. To a user, such a view appears like a single pattern language. The view V in Fig. 3, for example, appears to be a pattern language to create secure microservices in a cloud environment.

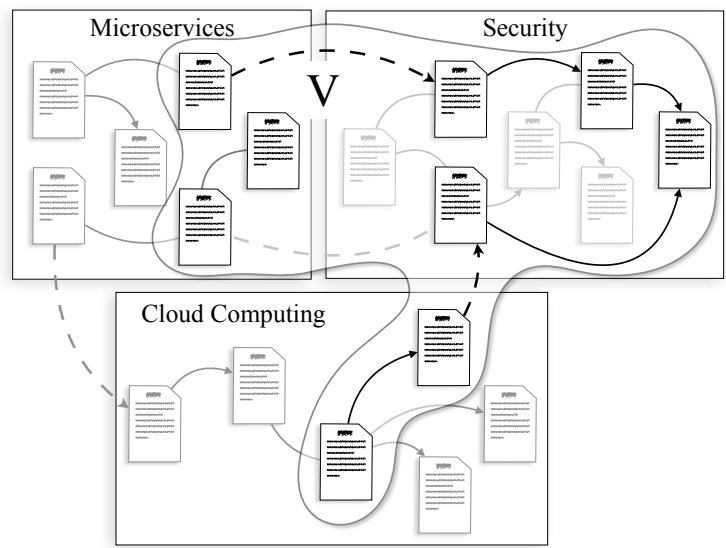

**Fig. 3**. A pattern language view.

Finally, it is far from trivial to determine "where to start" using a pattern language (be it a view or a basic pattern language). A user must understand all of the patterns, especially the problems each of the patterns solves, to find an *entry* to the pattern language [17], i.e. a first pattern where navigation through the pattern language begins and a solution path is created. By applying the solution path, some of the problems are solved, other remain. Based on the remaining problems another entry is determined and so on.

## 2.    Pattern Atlas

A tool that supports the creation of software architectures based on patterns has to cope with concepts like pattern languages, links, solutions, views, entries (and several more that we did not discuss). In this section we show by analogy how these concepts relate, and this analogy motivates the name of the tool.

### 2.1. PatternPedia

For more than a decade we built a tool called PatternPedia that supports several of the concepts above, and various pattern languages have been represented in PatternPedia [10]. The existing tool has been described in several publications, and it served as the basis for a couple of projects.

Also, repositories for working solutions (which are often domain specific, i.e. which cannot be generalized) have been created for some domains. Patterns from these domains point to corresponding working solutions. Concepts and tools to (semi-)automatically derive patterns from working solutions by means of data analytics have also been prototypically implemented.

Based on this experience it became clear that a new architecture and corresponding implementation for a general pattern repository (which encompasses solution repositories) is due: the *pattern atlas*. This name is justified because the realization of this new platform is steered by the analogy of an atlas that turned out to be helpful in decisions about the platform's functionality.

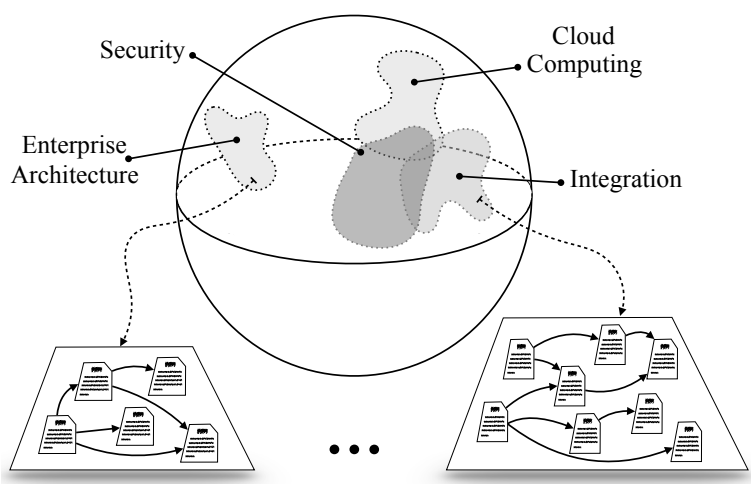

**Fig. 4**. Pattern languages as maps of IT domains.

### 2.2. Maps and Atlas: Covering IT with Pattern Languages

Information technology is a huge sphere, too huge to be able to present the knowledge about it as a whole with high precision. It is like the surface of earth, which needs a large collection of maps, i.e. an atlas, that represent information of various kinds to comprehend this surface.

Fig. 4 depicts the sphere of information technology as - well - a sphere, a mathematical sphere, i.e. as the surface of a ball in 3-dimensional space. The different domains of information technology are indicated by grey-shaded areas on this sphere. Examples of such domains are the domain of enterprise architecture that has been covered by a pattern language in [12], the domain of cloud computing in [11], the domain of microservices by [18], or enterprise application integration in [13]. In our analogy to earth, the grey-shaded areas are like geographic regions, e.g. countries, and the pattern languages correspond to maps of these regions. In cartography, a collection of maps that covers a certain part of earth is an atlas, and in analogy we call a collection of pattern languages covering a certain part of information technology a pattern atlas.

### 2.3.    Glueing Maps Together: Links Between Pattern Languages

In cartography, maps are flat representations of areas on earth, the latter of which are always curved (which becomes important a bit later). Differential geometry [19] considers the sphere, i.e. the surface of earth, as a 2-dimensional manifold and generalizes the notion of an atlas by emphasizing the functions that transform an area of the manifold into its flat image: an atlas is a set of pairs $\{(U_i, f_i)\}$, called charts, where $f_i : U_i \rightarrow V_i \subseteq \mathbb{R}^2$ is a "structure preserving" function. Then, $f_i(U_i)=V_i$ is what is known as a map in cartography. For $U_i \cap U_j \neq \varnothing$ the so-called transition function $f_j \circ f_i^{-1}$ determines how the maps $V_i$ and $V_j$ are glued together (see Fig. 5).

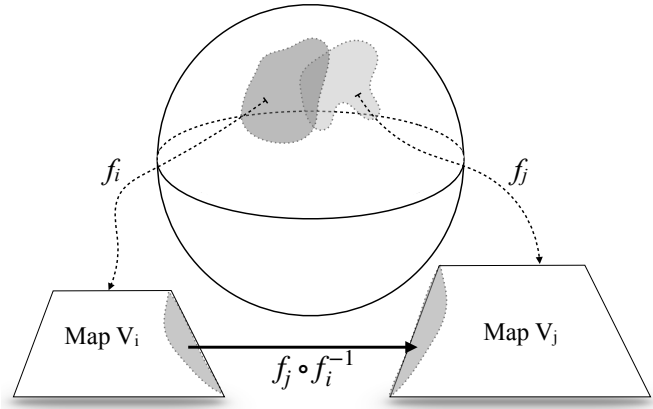

**Fig. 5.** Glueing maps together.

In our analogy, these transition functions are represented as links between pattern languages. The links define how various patterns of the affected pattern languages relate, i.e. how these pattern languages are glued together into a consistent map of the combined domain of information technology. For example, by establishing links between the microservice pattern language and the cloud computing pattern language a pattern language for microservices in the cloud is created.

### 2.4.    Special Representations of Maps: Views of Pattern Languages

An atlas in cartography supports various special representations of one and the same geographic region: special maps depict transport routes (e.g. highways or railroads), ecological zones (e.g. tropics, subtropics, deserts), mountain structures and so on.

Such representations are created by omitting details that are not relevant for the aspect of interest. These omissions allow to focus on specific aspects without having to understand or consider all the details of a geographic region.

The pattern atlas supports such omission to increase focus by means of views. A view is defined by selecting a subset of patterns and links of a given pattern language (or collection of linked pattern languages) to ease comprehension and to focus on certain aspects. For example, restricting the cloud computing patterns of [11] to the data related patterns eases the use of the pattern language for users that need to cope with data management in the cloud; or restricting the combined enterprise integration pattern language and security pattern language to corresponding communication-related patterns in both languages immediately supports users concerned with secured communication between applications.

### 2.5. Index for Finding Entities in Maps: Entries in Pattern Languages

When working with an atlas an index is used to efficiently find details within a geographic region like a certain city, a certain mountain, or a certain lake. Such an index is mainly a list of names of entities on earth and references to maps that contain their representations, as well as references to detailed positions within these maps. Finding a proper pattern to start solving a complex problem is far from being trivial, i.e. an analogy to an index is needed.

The concept of an entry point [17] corresponds to such an index, but due to the nature of patterns more sophistication is needed: an entry point is the starting pattern of a solution path solving a (probably composite) problem (note, that a solution path may consist of a single pattern only). It is determined based on the context of the overall architectural problem to be solved. A context is described by means of facts. "Negative facts" represent the problems to be solved. Since each pattern solves a problem, applying a pattern removes negatives facts, i.e. it turns them into "positive facts". The goal is to turn as many as possible negative facts into positive facts. Our proposed algorithms in [17] determines all possible solution paths addressing negative facts in the current situation, selects the solution path that will turn most negative facts into positive facts, and offers its start pattern as entry point to the pattern language. This assumes that the pattern language is extended by such facts, which is currently rarely the case.

Another approach to entry points that has been realized in PatternPedia is based on tags associated with patterns, or enabling full-text search on pattern documents: this way individual patterns can be determined that might help solving a problem. But this does not guarantee the determined pattern is the begin of a solution path that solved a maximum number of problems.

### 2.6. Concrete Renderings of Maps: Working Solutions

A map is a flat representation of a region on earth (see section 2.3). Ideally, the map should faithfully render the region. Here, faithful means that geometric properties like angles, areas, distances etc. of the region on earth are preserved, i.e. are (proportionally) the same on the map as in the region. But according the famous

Theorema Egregium by C.F. Gauss, any such flat rendering inherently results in distortions: i.e. a faithful rendering of a region on earth is impossible.

Important applications like navigation require maps that preserve at least one of the geometric properties of the region on earth. Luckily, this can be achieved. For example, equi-area projections preserve areas (e.g. Lambert's cylindrical projection). Azimuthal equidistant projections preserve distances (e.g. Postel's projection). Gnomonic projections preserve shorter routes. While such projections preserve one geometric property they distort the others: e.g. if areas are preserved, the shape of the areas is changed.

These concrete renderings of a map (i.e. such projections) serve specific needs like determining routes on sea, determining the distances between locations etc. While a map itself renders a region "as good as possible", it is not a proper solution to such specific problems. But the projections are "working solutions": a given map can be associated with a set of corresponding projections allowing to determine the shortest route between two locations on sea, to determine the area of a shape on the map etc. In this sense, maps are abstract while projections are concrete.

Similarly, the pattern atlas supports to associate concrete solutions (aka working solutions) with patterns. This way, the abstract solution sketched in the pattern document is made concrete. For example, if a pattern sketches how to use a message queue, concrete solutions specify how this is done in Amazon's SQS or IBM's MQSeries.

**Table 1.** Mapping concepts from patterns and cartography.

| Patterns | Cartography |
|---|---|
| Pattern Languages | Maps of geographic regions |
| Links | Arrangements of maps |
| Views | Special representations of regions |
| Solutions | Concrete renderings of a map |
| Entries | Index |

Table 1 shows how the discussed features and concepts from cartography correspond to the discussed features and concepts from pattern languages.

## 3.    Conclusion

We reminded the main concepts and entities from pattern languages, especially in their practical use in building software architectures. Extensions like views, entries, and working solutions have been summarized. Our prototypical support of pattern language-based design of software architectures named PatternPedia needs to be revamped after a decade of incremental development. The paradigm that guides us in this new implementation stems from cartography, which motivates the name of the new tool: Pattern Atlas. We have shown by analogy how the concepts and features of an atlas correspond to the concepts and features of the Pattern Atlas. The pattern atlas will be partially implemented in the project PlanQK, a platform to support quantum machine learning [23].

**Acknowledgements**

We are grateful to our colleagues Uwe Breitenbücher, Michael Falkenthal, Manuela Weigold and Karoline Wild for the discussions about the evolution of PatternPedia towards the Pattern Atlas.